\documentclass[amsmath,amssymb,amsbsy,prl,twocolumn,superscriptaddress,showpacs]{revtex4}
\usepackage[final]{showlabels}
\usepackage{amsfonts}
\usepackage{amsmath}
\usepackage{amssymb}
\usepackage{cancel}
\usepackage{bm}
\usepackage[usenames, dvipsnames]{color}
\usepackage{dcolumn}
\usepackage{epic}
\usepackage{epsfig}
\usepackage{eucal}
\usepackage{esdiff}
\usepackage{graphicx}
\usepackage[breaklinks,colorlinks = true,linkcolor = red,urlcolor=cyan,citecolor=red]{hyperref}
\usepackage{mathrsfs}
\usepackage{soul}
\usepackage{subfigure}
\usepackage{textcomp}
\usepackage{times}
\usepackage{wrapfig}
\usepackage{xy}

\def \veck {\mathbf{k}}
\newcommand{\dprime}{{\prime\prime}}
\newcommand{\eqn}[1] {eq.~(\ref{#1})}

\newcommand{\Fig}[1]{Fig.~\ref{#1}}
\newcommand{\Tab}[1]{Table~\ref{#1}}
\newcommand{\ourtitle}{Interplay of Magnetism and Topological Superconductivity  in Bilayer Kagome Metals}

\begin{document}

\title{\ourtitle}
\author{Santu Baidya}
\email{santubaidya2009@gmail.com}
\affiliation{Department of Physics and Astronomy, Rutgers, The State University of New Jersey, Piscataway, New Jersey 08854-8019, USA}

\author{Aabhaas Vineet Mallik}
\email{aabhaas.iiser@gmail.com}
\affiliation{International Centre for Theoretical Sciences, Tata Institute of Fundamental Research, Bengaluru 560 089, India}
\author{Subhro Bhattacharjee}
\email{subhro@icts.res.in}
\affiliation{International Centre for Theoretical Sciences, Tata Institute of Fundamental Research, Bengaluru 560 089, India}
\author{Tanusri Saha-Dasgupta}
\email{t.sahadasgupta@gmail.com}
\affiliation{Department of Condensed Matter Physics and Materials Science, S.~N. Bose National Centre for Basic Sciences, Kolkata 700098, India}

\date{\today}

\begin{abstract}
The binary intermetallic materials, $M_3$Sn$_2$ ($M$ = $3d$ transition metal) present a new class of strongly correlated systems that naturally allows for the interplay of magnetism and metallicity.  Using first principles calculations we confirm that bulk Fe$_3$Sn$_2$ is a ferromagnetic metal, and show that $M$ = Ni and Cu are paramagnetic metals with non-trivial band structures. Focusing on Fe$_3$Sn$_2$ to understand the effect of enhanced correlations in an experimentally relevant atomistically thin single kagome-bilayer, our {\it ab-initio} results show that dimensional confinement naturally exposes the flatness of band structure associated with the bilayer kagome geometry in a resultant ferromagnetic Chern metal. We use a multistage minimal modeling of the magnetic bands progressively closer to the Fermi energy. This effectively captures the physics of the Chern metal with a non-zero anomalous Hall response over a material relevant parameter regime along with a possible superconducting instability of the spin-polarised band resulting in a topological superconductor. 
\end{abstract}

\maketitle

\paragraph{Introduction :}  Accentuated quantum fluctuations due to dimensional confinement and  electron-electron correlations are at the heart of some of the novel electronic phases of condensed matter-- {\it e.g.}, high temperature superconductivity in cuprates\cite{cuprate} and iron pnictides\cite{pnic}; the low dimensional frustrated magnets\cite{frust}; easily exfoliable materials such as monolayer and bilayer graphene\cite{graphene, twisted}; and two dimensional (2D) electron gas leading to integer and fractional quantum Hall effects\cite{hetero} in synthetic heterostructures\cite{synthetic1,synthetic2}.

A recent addition to this ongoing research is the binary intermetallic series $M_{m}$$X_{n}$ with $M\ =\ 3d$ transition metals (TMs) forming stacked kagome layers, separated by $X$ = Sn, Ge spacer layers in stoichiometric ratios of $m$:$n$ = 3:1, 3:2 or 1:1.  Recent experiments report the observation of bulk Dirac cones in the electron band-structure \cite{dirac-fe,fe3sn2}, large anomalous Hall response \cite{Nakatsuji15,Nakatsuji16,mn3ge,mn3sn} as well as  magnetic Weyl excitations\cite{weyl} in them.  Diverging density of states, probed by scanning tunneling spectroscopy, have been reported for Fe$_3$Sn$_2$\cite{fe3sn2-sts} as well as for the ternary kagome ferromagnetic compound, Co$_3$Sn$_2$S$_2$\cite{cosns-sts}, though the existence of flat band, a characteristic to kagome geometry, has remained inconclusive possibly due to hybridisation with other bands in the 3D material. Recently synthesised,\cite{fesn} bulk FeSn has been claimed to have flat bands at energies few hundreds of meV below and above Fermi level. In this backdrop, it is curious to explore the consequence of dimensional confinement in these intermetallics by considering the atomistically thin limit of these materials which is expected to further enhance the strong correlation physics and thereby providing a platform for the interplay of flat-band physics and fluctuating magnetism in the low-dimensional itinerant systems containing $3d$-TMs.

\begin{figure}
\includegraphics[scale=0.35]{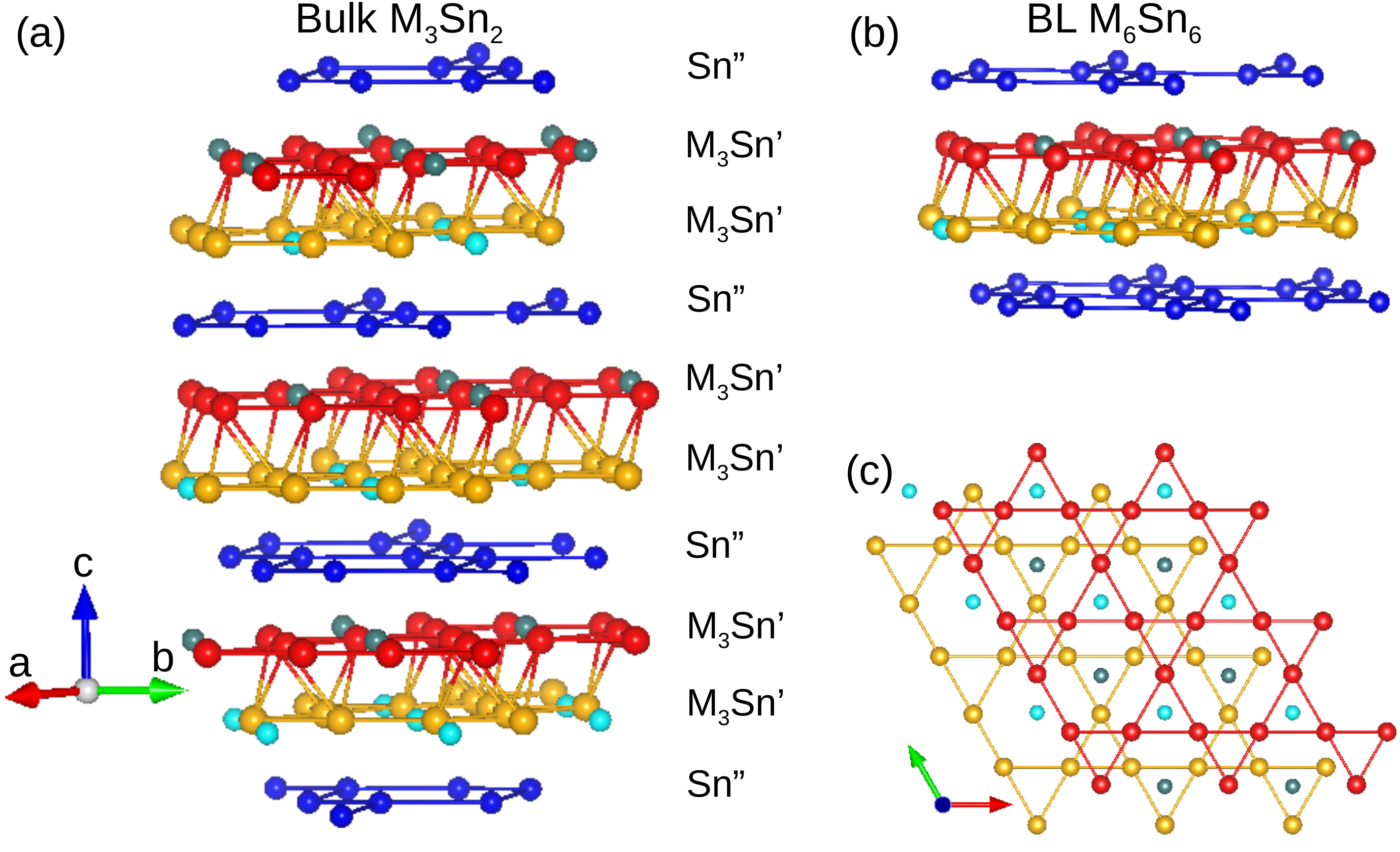}
\caption{(a) Layered arrangement in bulk $M_3$Sn$_2$. (b) Bilayer $M_6$Sn$_6$ derived out of the bulk layered structure. (c) Stacking of two kagome layers within the bilayer, viewed along the out-of-plane direction.}
\label{fig:lat}
\end{figure}

In this Letter, we explore the above possibility within the framework of {\it ab-initio} density functional theory (DFT) and effective low energy minimal models inspired by the DFT band structure.   To probe the effect of dimensional confinement, we consider one unit of kagome bilayer derived from the bulk structure (Fig.~\ref{fig:lat}(a)), sandwiched between two Sn layers (Fig.~\ref{fig:lat}(b)). Also, in addition to Fe compound which is already synthesized as a bulk material, we consider two more late TM based compounds -- Ni and Cu-- which are yet-to-be synthesized, in order to understand the generic behaviour of this family of materials. The calculated cleavage energy\cite{benedek,santos} costs involved in creation of a bilayer is $1-2$ J/m$^{2}$ \cite{sup}-- similar to that required for creating 2D MXenes, the 2D counterparts of MAX phases\cite{mxene,MAX}, which have already been successfully synthesised through chemical etching.

Analysis of the electronic structure of bulk $M_3$Sn$_2$ ($M$ = Fe, Ni, and Cu) prompts us to conclude that Fe$_3$Sn$_2$ is the natural choice to search for fluctuation driven physics in magnetic flat bands, as both Ni$_3$Sn$_2$ and Cu$_3$Sn$_2$ turn out to be non-magnetic within our DFT calculations. Thus, while Ni$_3$Sn$_2$ and Cu$_3$Sn$_2$ have interesting band structures (see below) and hence deserve attention, Fe appears to be in a {\it sweet spot} of the interplay of correlations and band physics in the late $3d$ TM series. Our DFT calculations for bilayer Fe$_3$Sn$_2$ reveal that confinement to the bilayer limit, results in the formation of near-flat bands within $\pm$ 10 meV of Fermi level, together with massive Weyl-like band features. Interestingly, the ferromagnetic correlations in the 3D compound are found to survive down to bilayer limit. The resulting almost flat bands have non-zero Chern number, realising a Chern metal in the bilayer system. Inclusion of magnetic fluctuation effects within the model calculations of the kagome bilayer lead to a fluctuation-driven topological superconductor.

\paragraph*{\it Electronic Structure of Bulk $M_3$Sn$_2$ --}  Rhombohedrally structured bulk $M_3$Sn$_2$ consist of kagome bilayers of $M$ atoms sandwiched between stanene layers (Sn${''}$ atoms) as shown in Fig. \ref{fig:lat}(a) \cite{fe3sn2-struc}. The crystal structure\cite{sup} allows for a breathing anisotropy in each kagome plane resulting in unequal sizes of the up and down triangles in the kagome planes (Fig.~\ref{fig:lat}(c)). The crystal structures of Ni$_3$Sn$_2$ and Cu$_3$Sn$_2$ are obtained from that of Fe$_3$Sn$_2$ through symmetry-allowed relaxation.

\begin{figure}
\includegraphics[scale=0.315]{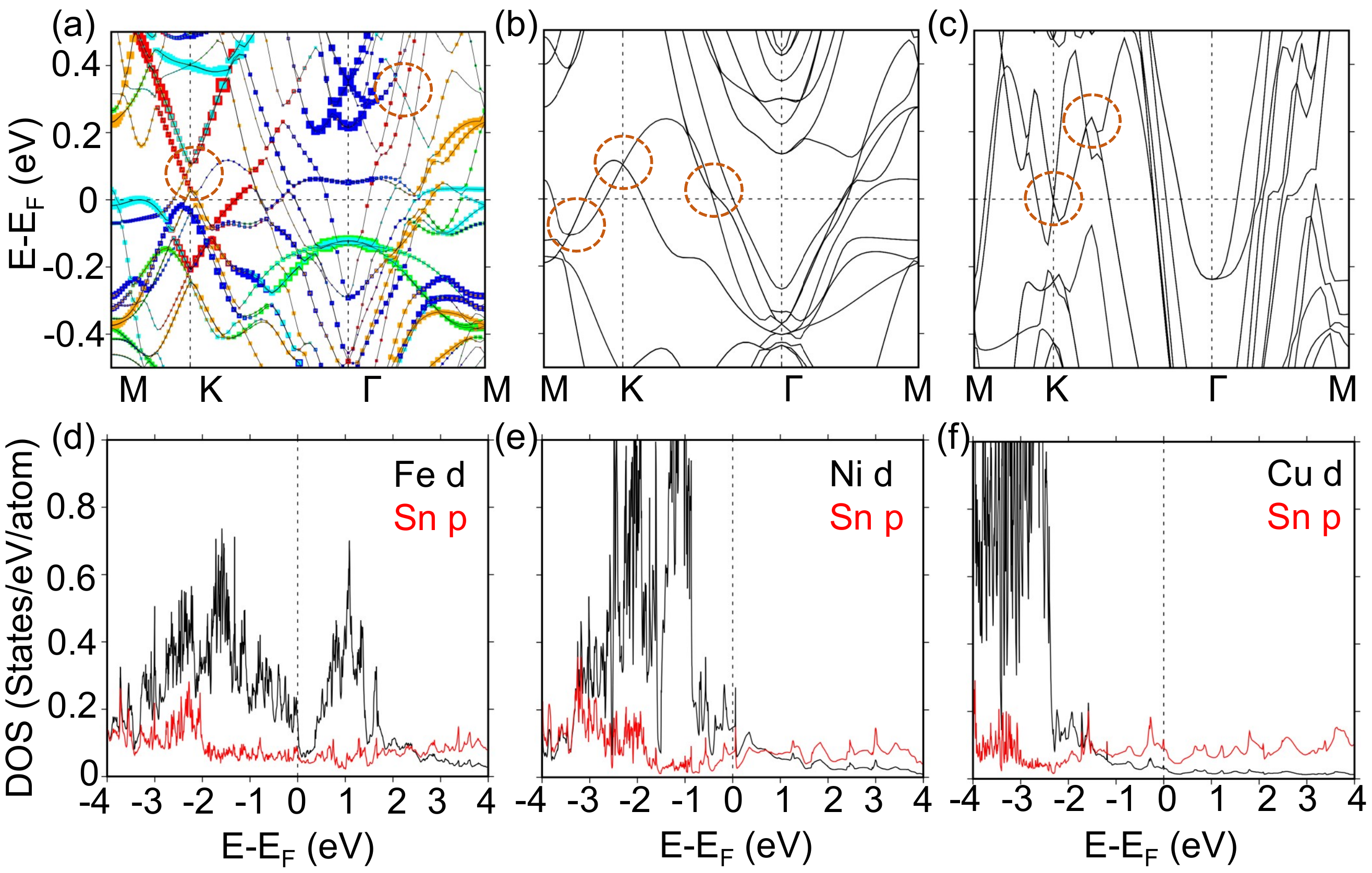}
\caption{GGA+$U$+SOC band structure of bulk ferromagnetic Fe$_3$Sn$_2$ (a), paramagnetic Ni$_3$Sn$_2$ (b) and Cu$_3$Sn$_2$ (c). The non-trivial Weyl and Dirac-like crossings are encircled. In (a) the Fe $d$-orbital character are shown-- red : $d_{xy}$, green : $d_{yz}$, cyan : $d_{xz}$, blue : $d_{3z^2-r^2}$, yellow : $d_{x^2-y^2}$. The corresponding non-spin-polarized projected DOS are shown for the Fe (d), Ni (e) and Cu (f) compounds. }
 \label{fig:bulk_DFT}
\end{figure}

 The DFT calculations were performed in plane wave basis using Vienna Ab-initio Simulation Package\cite{PAW,VASP} with exchange-correlation functional within generalised gradient approximation (GGA)\cite{PBE}. Also,  correlation effect at TM sites is included within GGA+$U$\cite{Liechtenstein95} with $U$ = 0.5 eV \cite{fe3sn2-arpes}. Weak spin-orbit coupling (SOC) is also included for the TM 3$d$ states which appears to be a crucial ingredient to drive the topological behaviour as well as the stability of magnetism in the bilayer limit (see below). Further details are provided in the Supplementary Material (SM)\cite{sup}. Figs.~\ref{fig:bulk_DFT}(a)-(c) show the GGA+$U$+SOC band structure of Fe$_3$Sn$_2$, Ni$_3$Sn$_2$ and Cu$_3$Sn$_2$ plotted along the high symmetry directions of the hexagonal Brillouin zone (BZ) \cite{footnote1}.

Interestingly, while the DFT calculations for Fe$_3$Sn$_2$ stabilises a ferromagnet at the Fe sites with moment $\approx 2.2~\mu_B$, in agreement with reported experiments\cite{fe3sn2-arpes}, both Ni and Cu compounds turned out to be paramagnetic. Focusing on the corresponding density of states (DOS) (Figs~\ref{fig:bulk_DFT}(d)-(f)), we find that while in case of Fe$_3$Sn$_2$, the low-energy states are primarily dominated by Fe $d$ states with small admixture from Sn $p$ due the covalency, for  Ni$_3$Sn$_2$ and Cu$_3$Sn$_2$ there is progressively higher contribution of the Sn-$p$ orbitals such that for Cu it is almost entirely of Sn $p$ character. The Stoner criteria of magnetism, appropriate for metallic systems, gives $ I \times N(E_F)$ ($I$ = Stoner parameter, $N(E_F)$ = DOS at $E_F$) to be larger than one (1.4) for Fe$_3$Sn$_2$, and significantly less than 1 for Ni$_3$Sn$_2$ and Cu$_3$Sn$_2$ (0.3 and 0.1, respectively), justifying the ferromagnetic (paramagnetic) ground state in the Fe (Ni and Cu) system(s).  The band structures of Fe$_3$Sn$_2$, which is in good agreement with literature \cite{fe3sn2-arpes}, and that of Ni$_3$Sn$_2$ and Cu$_3$Sn$_2$ show topologically non-trivial Weyl/Dirac-points (Fig.~\ref{fig:bulk_DFT}(a-c)). The electronic structure of Ni$_3$Sn$_2$ and Cu$_3$Sn$_2$ compounds though appear interesting and deserves further attention, in keeping the focus on the interplay of magnetic fluctuations, topological properties and low dimensionality, we discuss the properties of bilayers of Fe$_3$Sn$_2$ in the rest of this Letter.

\begin{figure}
\includegraphics[width=0.475\textwidth,keepaspectratio]{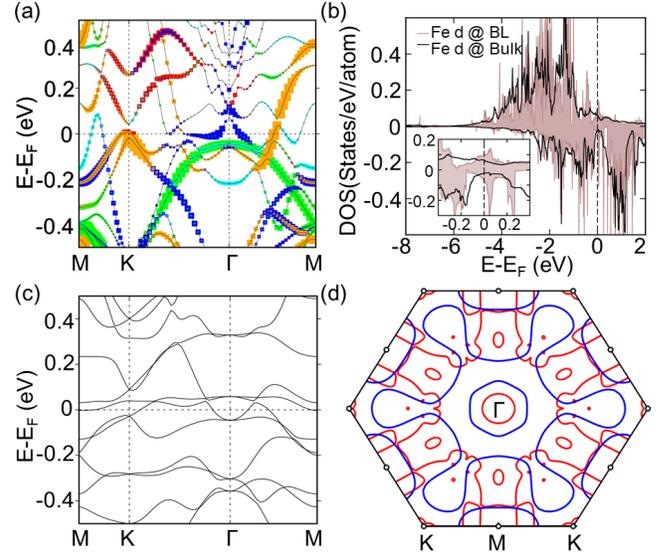}
\caption{GGA+SOC+$U$ band structure (a) and Fe $d$ projected spin-polarized DOS, with positive (negative) axis corresponding to DOS in up (down) spin channel (b) of  bilayer Fe compound in the FM state. In (a) the Fe $d$ orbital characters are denoted with different colors-- red: $d_{xy}$, green: $d_{yz}$, cyan: $d_{xz}$, blue: $d_{3z^{2}-r^{2}}$ and
  yellow: $d_{x^{2}-y^{2}}$. (b) shows comparison between the bilayer and  bulk DOS, while the inset shows the van Hove features of the two-dimensional electronic structure of
  the bilayer. (c) Band structure of bilayer in nonmagnetic state. (d) the Fermi surface in the non-magnetic (blue) and ferromagnetic states (red).}
\label{fig:BL_DFT}
\end{figure}

\paragraph{Electronic Structure of Bilayer FeSn : } The Sn$''$ terminated Fe$_6$Sn$_6$ bilayer is shown in Fig.~\ref{fig:lat}(b). The kagome bilayer, which in this case is isolated, consists of two kagome layers of Fe atoms, shifted with respect to each other. It is important to note that the bulk FeSn structure studied recently\cite{fesn} consists of alternate Fe-kagome layers and Sn layers and hence is different from the present case. 

The GGA+SOC+$U$ band structure of the Sn-terminated bilayer FeSn (cf Fig.~\ref{fig:lat}(b)) is shown in Fig.~\ref{fig:BL_DFT} along with the orbital characters of the bands projected to the Fe $d$. DFT estimate for intra and inter kagome layer magnetic interactions for the above bilayer  are both ferromagnetic-- as in bulk Fe$_3$Sn$_2$-- with magnitudes $\approx 10$ meV and $\approx 0.3$ meV respectively. This is in contrast to magnetic behaviour of bulk FeSn reported recently\cite{fesn}, where the ferromagnetic Fe kagome layers are coupled antiferromagnetically. 

The bilayer band structure shown in Fig. \ref{fig:BL_DFT}(a) should be contrasted with the bulk electronic band structure (\Fig{fig:bulk_DFT}(a)). The latter is characterized by several low energy bands having non-Fe character as expected for a three dimensional network of the Sn atoms in the the bulk. In comparison, the band structure of bilayer FeSn within $\pm$ 10 meV is primarily of Fe $d_{x^{2}-y^{2}}$, $d_{xy}$ and $d_{3z^{2}-r^{2}}$ characters, with little contribution from Sn $p$. While the basic features of the bilayer DOS, presented in Fig.~\ref{fig:BL_DFT}(b), is similar to that of bulk, close observation reveals (inset of \Fig{fig:BL_DFT}(b)) formation of van Hove like singularities in the DOS of the bilayer arising from dimensional confinement. This is further confirmed by calculating effective masses of the low-energy bands. While the parallel components of effective masses range between $3-5~m_e$, that in perpendicular direction is $1000-2000~m_e$.

Most importantly, in contrast to the bulk band structure, the bilayer band structure shows a couple of somewhat {\it flat bands,} arising due to destructive interference of hoppings in the kagome bilayers, in an energy window of $\pm 10$ meV around the Fermi level, $E_F$, in a forced paramagnetic state (Fig.~\ref{fig:BL_DFT}(c)). They survive, albeit somewhat dispersive, in the magnetic state (Fig.~\ref{fig:BL_DFT}(a)). The Fermi surface of the paramagnetic and ferromagnetic states are shown in Fig.~\ref{fig:BL_DFT}(d). The reconstruction of the Fermi surface due to ferromagnetic order resulting in small electron and hole like pockets is apparent. 

The nearly flat bands span a significant portion of the BZ and should be contrasted with  the case of bulk FeSn structure\cite{fesn} consisting of kagome monolayers, where the flat band features occur at few hundred meV from $E_F$.  Due to ferromagnetic order, time-reversal symmetry is broken and these flat-ish bands acquire a finite Chern number. Constructing the maximally-localized Wannier functions using WANNIER90\cite{wannier90}, we calculated the integrated Berry curvature over the 2D BZ. This gives a  Chern number of -1 for the flat band closest to $E_F$. Thus the DFT results show that bilayer FeSn may stabilize a ferromagnetic Chern metal\cite{chernm1,chernm2} with  non-quantized but large anomalous Hall response.

\paragraph{Effective tight binding model :} Based on our DFT findings, we conclude that geometric confinement to bilayer results in (a) quasi-2D electronic structure, (b) survival of ferromagnetic correlation, and (c) realisation of almost flat bands. The effect of fluctuations on almost flat bands are expected to be strong, opening up possibilities for stabilising novel phases.  The intricate features of the low energy DFT bands near $E_F$ (cf Fig.~\ref{fig:BL_DFT}(a)) require a detailed tight-binding model accounting for the various hopping processes involved. Here instead,  we construct simpler tight-binding models with the right orbital character and short range hopping that capture qualitatively the low energy DFT band structure and use them to study the effect of correlation and band properties. 

\begin{figure}
	\includegraphics[width=1.0\linewidth]{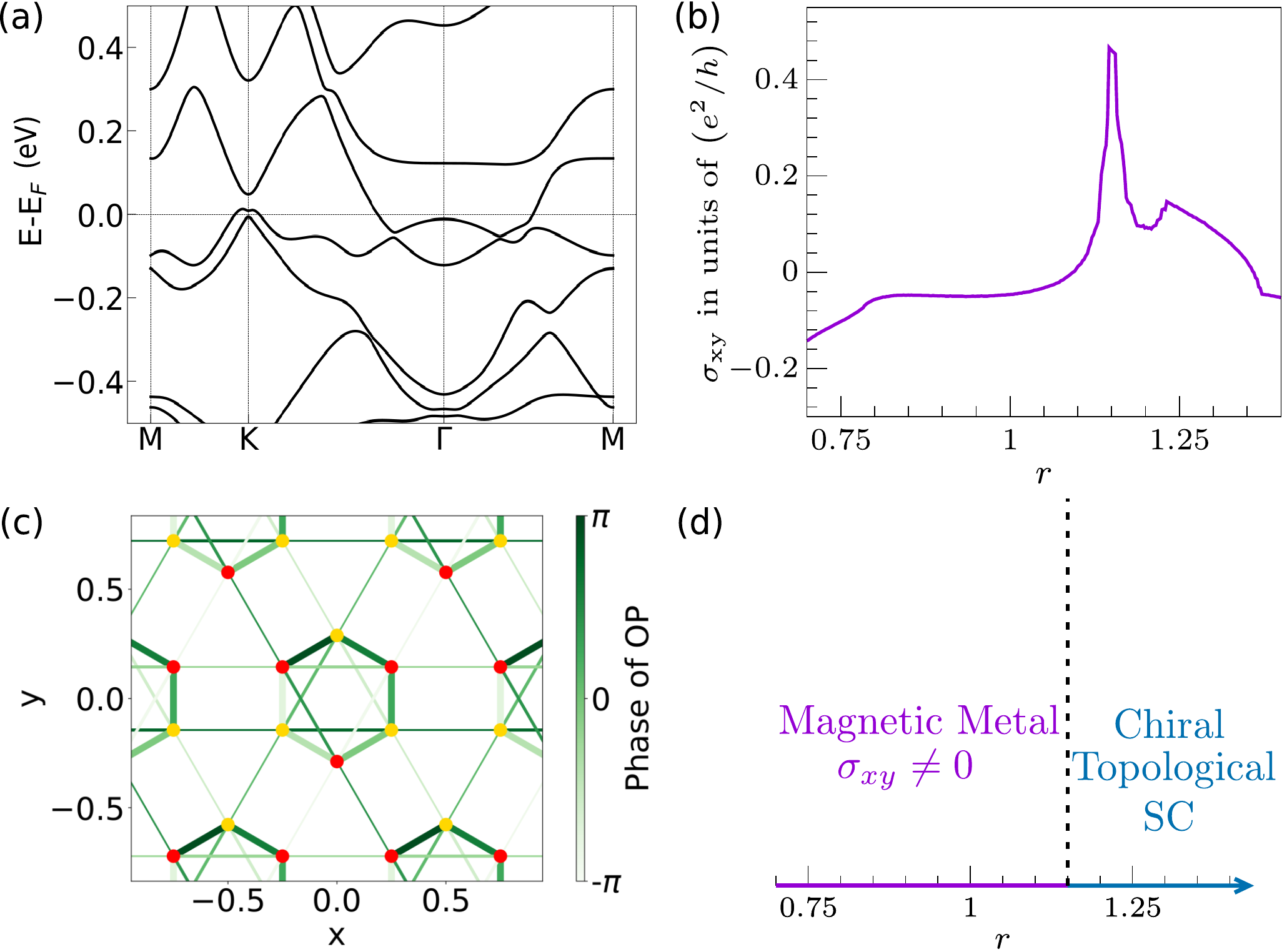}
	\caption{Panel (a) shows tight binding bands obtained for a three orbital model with parameters tuned to reproduce qualitative agreement with the DFT results. Panel (b) shows the evolution of $\sigma_{xy}$ as a function of the breathing anisotropy parameter $r$ (see text). Panel (c) shows the superconducting order parameter in the reduced one orbital spin polarised tight-binding model  with nearest neighbour attractive interactions mediated by ferromagnetic fluctuations. The thickness of the bonds is proportional to the amplitude of the superconducting order parameter and the colors encode its phase. Panel (d) shows the topological superconductor and magnetic metal (with $\sigma_{xy} \neq 0$) as a function of $r$ for the one-orbital tight-binding model.}
	\label{fig:tb_bands}
\end{figure}

To this end we introduce two related symmetry allowed models-- (1) a three orbital (plus spin)/site model in which we account for the magnetisation within mean field decomposition of onsite Hubbard interactions in the ferromagnetic channel. This captures the large anomalous Hall response in the Chern metal phase (Fig.~\ref{fig:tb_bands}(b)) in a material relevant parameter regime; and, (2) an even more simplified one spin polarized orbital/site tight-binding model which captures the flatish band near the Fermi level which we use to study the possible superconductivity driven by magnetic fluctuations within a self consistent Bardeen-Cooper-Schrieffer (BCS) mean field theory. We expect that since the superconductivity arises from the instability of the Fermi surface the minimal one orbital model gives a valid qualitative description of such phases. The parameters of both the models  reveal that while the nearest neighbour hopping within each kagome layer dominates\cite{sup}, further neighbour and interlayer hoppings are non-negligible-- indicating differences in the nature of the almost flatbands in bilayers from that in a single-layer kagome.

Our minimal tight binding modeling starts by \cite{sup,Xu15} including three Fe $d$-orbitals per site-- $d_{3z^2-r^2}$, $d_{x^2-y^2}$ and $d_{xy}$-- which contribute to the electronic states within $\pm$ 10 meV of the DFT band structure (cf \Fig{fig:BL_DFT}(a)). This generic symmetry allowed model allows intra and inter kagome layer first, second and third nearest neighbour hopping among the orbitals, along with (weak) SOC projected to the above orbitals.  As noted above, the breathing anisotropy within each kagome layer generically leads to difference in the hopping amplitude on bigger-triangles compared to smaller-triangles. We introduce a parameter $r$ in the tight-binding model as the ratio of the nearest neighbour hopping amplitudes on bigger-triangles to that on smaller-triangles. Generically, we do not know the value of $r$ in bilayer FeSn and related materials. Further it is possible that the actual value may depend on synthesis process and substrate and therefore is an experimentally relevant parameter which we vary to study various properties. Tuning the parameters of this model \cite{sup} lead to a representative  tight-binding band structure in \Fig{fig:tb_bands}(a)) with $r=1.25$ which is in semi-quantitative agreement with the DFT results (\Fig{fig:BL_DFT}(a)).  

\paragraph{Chern Metal and Anomalous Hall effect :} As suggested by our WANNIER90 Chern number calculation mentioned above, the bands close to the Fermi level in \Fig{fig:tb_bands}(a) also have non-zero Chern numbers. While it is complicated to calculate the Chern numbers of the individual bands since they cross, a more robust quantity is the anomalous Hall response of the resultant Chern metal. We plot the anomalous Hall conductivity ($\sigma_{xy}$) as a function of the asymmetry parameter $r$ in Fig.~\ref{fig:tb_bands}(b) where other band parameters are kept fixed. The resultant finite response over a wide parameter regime $r\in (0.7-1.4)$ indicates that the  anomalous Hall response is a robust feature of Chern metal in bilayer FeSn and allied materials.

The Chern metal is an extremely interesting phase where the partially filled band has a topological invariant. The instability of such a metallic phase therefore involves an intricate interplay of band topology and correlations. In presence of small SOC, we expect the ferromagnetic order to be stable at finite temperatures even in the bilayer, albeit with enhanced magnetic fluctuations. These magnetic fluctuations can then act as a pairing glue leading to superconductivity in the spin-polarised band. We now focus on the possibility of realising such a magnetic fluctuation driven superconductor.

\paragraph*{Superconductivity:} Both the DFT and the tight-binding model show the presence of small Fermi pockets near the $K$ and the $\Gamma$ points of the BZ arising from the almost flat low energy bands (Figs. \ref{fig:BL_DFT}(d) and \ref{fig:tb_bands}(a)).  Magnetic fluctuations lead to effective attractive interactions for electrons-- driving a superconducting instability naturally in the triplet channel for the spin-polarised bands\cite{Berk66,Doniach66,Samokhin08,Wang16,Wu17}. This superconductivity can be explored within a self-consistent BCS mean field theory. For this purpose we use a symmetry allowed effective tight binding model with one spin polarized orbital/site of the kagome bilayer, with parameters chosen such that the DFT bands close to $E_F$  are well represented \cite{sup}. Further, integrating out the magnetic fluctuations leads to short-range (nearest neighbour in our case) attractive interactions, $V$, between the electronic densities, $\sim -V \sum_{\langle ij\rangle} n_in_j$. Within a multi-band BCS mean field theory, the above model indeed stabilises a superconductor for a wide regime of the parameter $r$ when $V\sim t$\cite{sup}, where $t$ is the nearest neighbour hopping of the effective low energy one-orbital tight-binding model.  The fairly large value of the interaction can be attributed to the small density of states at $E_F$ due to the small Fermi-pockets. Here we plot the results for the representative choice of $V=2t$. A rough estimate of effective low energy scales gives $t\sim 0.13$ eV and hence $V\sim 0.26$ eV $\approx 0.5 U$\cite{sup}.

In our analysis, we incorporate eighteen pairing order parameters corresponding to the nearest neighbour bonds associated with a unit cell \cite{sup}. In \Fig{fig:tb_bands}(c) we show the superconducting order with the thickness of a bond being proportional its magnitude and the color of the bond encoding its phase. The maximum pairing amplitude is $\sim 0.02t$ which would correspond to a mean field transition temperature of $\sim 10K$. A straightforward analysis reveals that the pairing amplitudes in \Fig{fig:tb_bands}(c) transform like a $l_z = 1$ orbital under a rotation by $2\pi/3$ about the centre of the hexagon of the bilayer akin to $p_x+ip_y$ superconductor\cite{Wang16}. The topological nature of superconductivity in this system is easily confirmed by computing the net Chern number of the negative energy Bogoliubov bands \cite{Qi10,Tanaka12,Liu17,Fukui05}. Remarkably, this topological superconductor is stable over a wide range of the breathing anisotropy parameter, $r$ which is promising in regard to its experimental detection. For sufficiently small values of $r$ superconductivity ceases to exist and one recovers a magnetic metal exhibiting anomalous Hall response. This is shown in \Fig{fig:tb_bands}(d).

\paragraph{Summary and Outlook :} Our DFT results show that kagome intermetallic series derived from bulk $M_3$Sn$_2$ ($M$ = Fe, Ni, Cu) can host a rich interplay of band physics and correlations. To the best of our knowledge, while only bulk Fe$_3$Sn$_2$ has been synthesized, the Ni and Cu counterparts provide future avenues to explore. The above interplay is most prominent in the case of Fe where dimensional confinement in the bilayer limit enhances it by stabilising a ferromagnetic metal with nearly flat bands near the Fermi level and thereby giving a Chern metal with large anomalous Hall conductivity. Instability of this Chern metal, within a low energy tight-binding model and BCS mean field theory results in a topological superconductor in a material relevant parameter regime. A related instability, particularly relevant for the nearly flat band Chern metal, is a magnetic fluctuation driven fractional Chern insulator. It would be interesting to investigate the relevance of such a novel phase in the present context. All the above ingredients have close similarity with the rich physics of twisted bilayer graphene and hence experimental progress in isolating bilayer Fe$_3$Sn$_2$ and related materials may open up newer playgrounds of novel correlated physics probing the interplay of band topology, electron-electron correlations and spontaneous symmetry breaking.

\acknowledgements
The authors thank H. R. Krishnamurthy, V. B. Shenoy, S. Nakatsuji, J. Checkelsky, M. Jain and A. Agarwala for useful discussions. The authors thank Yogesh Singh for a careful reading of the manuscript. TSD and SBh thank JNU, New Delhi; ICTS-TIFR, Bangalore and SBh thanks ISSP, Tokyo for hospitality during different stages of this project.  AVM and SBh acknowledge financial support through Max Planck partner group on strongly correlated systems at ICTS; SERB-DST (Govt. of India) early career research grant (No. ECR/2017/000504) and the Department of Atomic Energy, Govt. of India, under Project no.12-R\&D- TFR-5.10-1100.\\

S. B. and A. V. M. have contributed equally to this work.

\newpage

\begin{center}
{\bf \Large {Supplementary Information}}
\end{center}

\section{DFT Computational Details}

The first principles electronic structure calculations have been carried out
within the framework of DFT in the plane wave basis with projector augmented wave (PAW)
potential\cite{PAW} and generalized gradient approximation (GGA) of Perdew-Burke-Ernzerhof (PBE)\cite{PBE} for exchange-correlation as implemented in Vienna Ab initio Simulation Package (VASP).\cite{VASP}  A k-point grid of 8$\times$8$\times$2, used to discretize the first Brillouin
zone and a plane wave cut-off of 600 eV were found to be sufficient to achieve convergence
of total energy with respect to k-point and energy cut-off. The energy convergence criterion
was set to 10$^{- 8}$ eV during the energy minimization process of the self-consistent cycle.

Starting from the experimental structure, \cite{fe3sn2} full optimization of the Fe$_{3}$Sn$_{2}$ crystal structure was carried out with the force criterion of 10$^{-3}$ eV/$\AA$. The crystal structures of yet-to-be synthesized Ni$_{3}$Sn$_{2}$ and Cu$_{3}$Sn$_{2}$ were obtained by starting from the experimental structure of Fe$_{3}$Sn$_{2}$, and replacing Fe by Ni and Cu, respectively, followed by full relaxation including the cell and the atomic positions.
The bilayer structures were simulated starting from the fully optimized respective
bulk $M_{3}$Sn$_{2}$ ($M$ =Fe, Ni and Cu) structures, terminated at desired layers on both side
within a slab geometry. Within the periodic set-up of the calculations, the bilayer structures
were separated by a sufficiently large vacuum layer of 10 $\AA$ to ensure minimal interaction
between consecutive bilayers. The atomic positions in the bilayer geometries
were relaxed keeping the in-plane lattice constant fixed to the value of the relaxed
bulk counterparts. The electronic structure calculations for the bilayer compounds were carried out using k-point grid of 12$\times$12$\times$1.      

Due to the presence of correlated metal cations Fe, Ni and Cu, a supplemented Hubbard $U$ correction of 0.5 eV was considered at the metal sites within the framework of GGA+$U$ calculations. The
choice of $U$ value of 0.5 eV was made following that in recent literature.\cite{fe3sn2-arpes}
The ground state electronic structure presented in the manuscript were calculated considering both the Hubbard $U$ and the spin-orbit coupling effect using GGA+$U$+SOC method.

\begin{figure}
\centering
\includegraphics[width=0.45\textwidth,keepaspectratio]{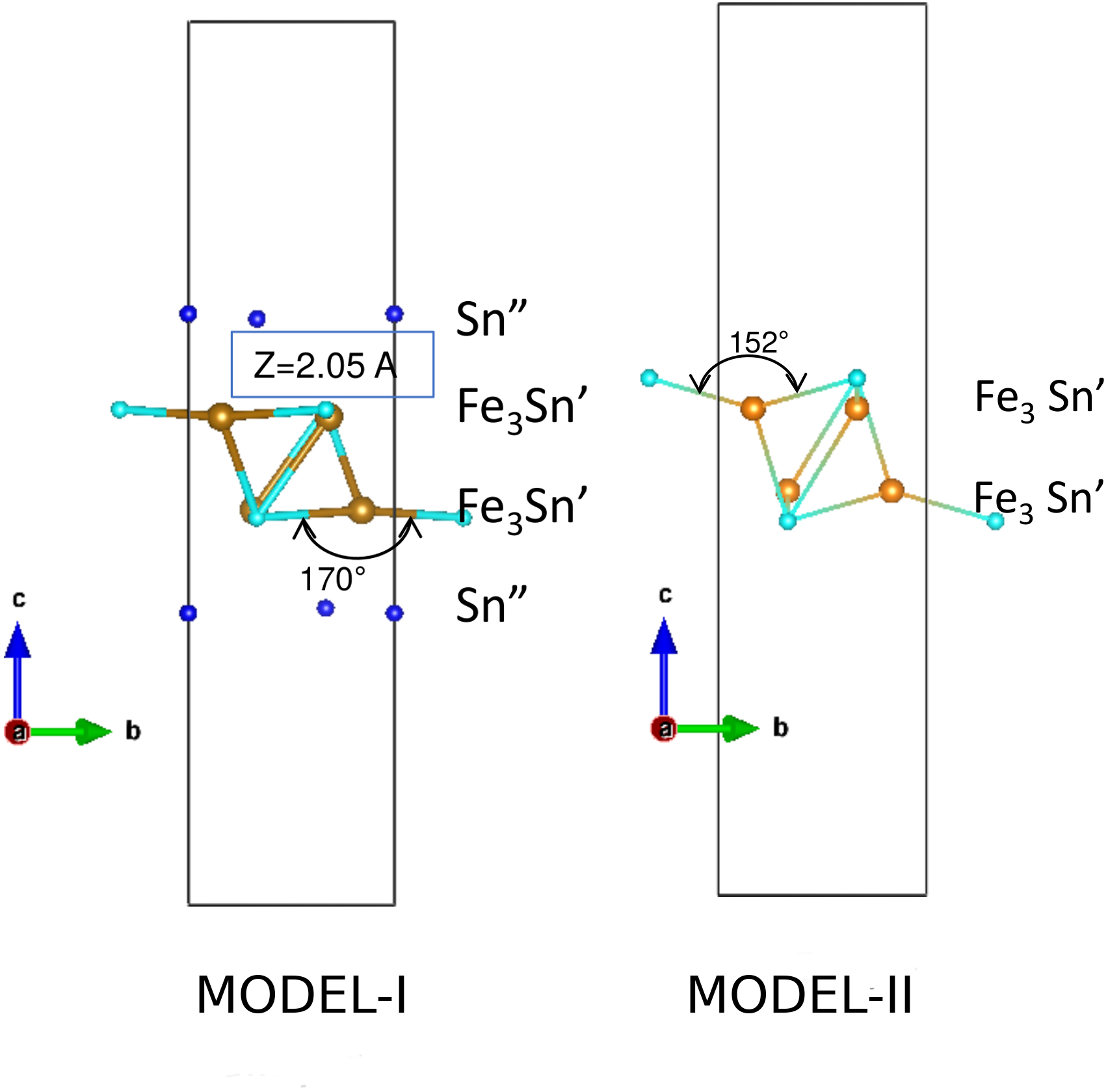}
\caption{Two possible bilayer geometries, derived out of $M_3$Sn$_2$ structures,
Model-I (left) and Model-II (right).}
\label{fig:lattice}
\end{figure}

\subsection{Cleavage Energy}

Possible means to derive the 2D counterparts from the 3D layered structures
are through mechanical or chemical route. An energy barrier needs to be overcome,
in this respect, known as the cleavage energy.
The cleavage energy ($E_{cl}$) can be defined as the
energy required to generate two (top and bottom) surfaces by cleaving the bulk
layered structure along the desired plane,\cite{benedek,santos}
$E_{cl} = 2\times \dfrac{(E_{slab} -E_{bulk})}{2A}$
where $E_{slab}$ is the total energy of the cleaved system with two exposed surfaces and $E_{bulk}$ is the
total energy of the same in bulk configuration, $A$ being the surface area.

Depending on the nature of the terminating surface, the atomistically thin layer
limit of the studied materials can be obtained from the bulk structure in two different ways keeping intact the kagome layers, one by terminating at Sn$''$ layer (referred as Model-I) or at Fe-Sn$'$ layer
(refereed as Model-II), as shown in Fig. S1. The chemical formula of Model-I is $M$Sn, while that of
Model-II is $M_3$Sn. We calculated cleavage energies of both the models for Fe, Ni and Cu systems.
For this purpose, the separations between the slabs were gradually increased in the
range 0.2-10 $\AA$ and the surface area-normalized total energy differences between the slab-separated
and bulk models were calculated, which show saturation when the separation exceeded 6 $\AA$, the value at
saturation being the cleavage energy. \Tab{tab:cleav_ener} shows the computed values of cleavage
energies for all three compounds both in Model-I and Model-II.

\begin{table}[t]
  \begin{tabular}{|c|c|c|c|}
    \hline
    Model & Fe & Ni & Cu \\ \hline
    Model-I & 1.762 & 2.083 & 1.105 \\
    Model-II & 1.923 & 2.267 & 1.576 \\\hline
  \end{tabular}
  \caption{Calculated cleavage energies in J/m$^{2}$ for Model-I and Model-II of Fe, Ni and Cu compounds.}
  \label{tab:cleav_ener}
  \end{table}

  Our calculation of the cleavage energies showed that the termination at the Sn${''}$ layer to be somewhat favorable over that at Fe-Sn$'$ layer, as found in recent study.\cite{fesn} In the text we thus focused only
  on Model-I.

\subsection{Crystal Structures}  
\label{sec:xtal_struct}
\begin{figure*}
\centering
\includegraphics[width=1.0\textwidth,keepaspectratio]{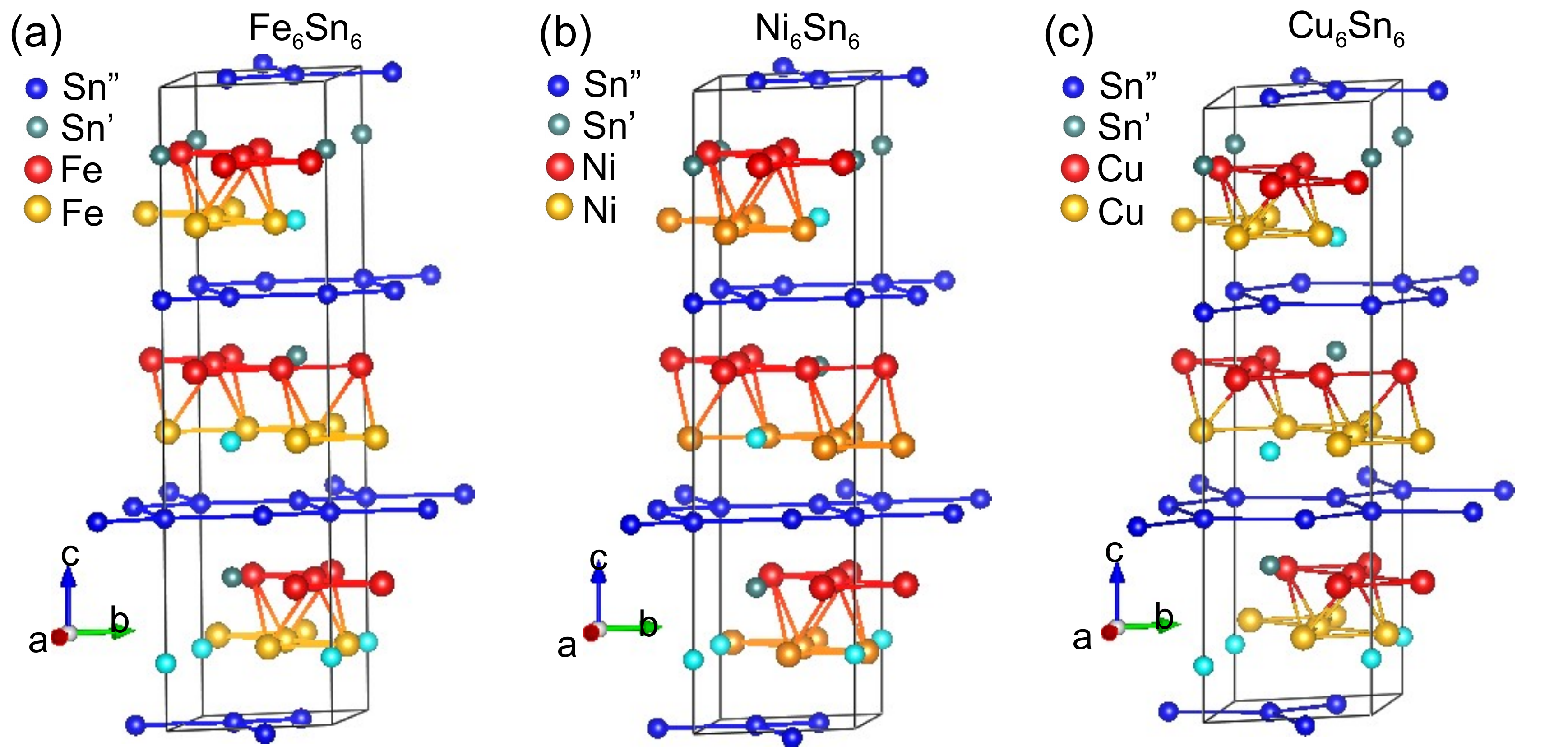}
\caption{Structure of bulk (a) Fe$_6$Sn$_6$ , (b) Ni$_6$Sn$_6$ and (c) Cu$_6$Sn$_6$ compounds.}
\label{fig:bulk_struct}
\end{figure*}

\begin{figure*}
\centering
\includegraphics[width=1.0\textwidth,keepaspectratio]{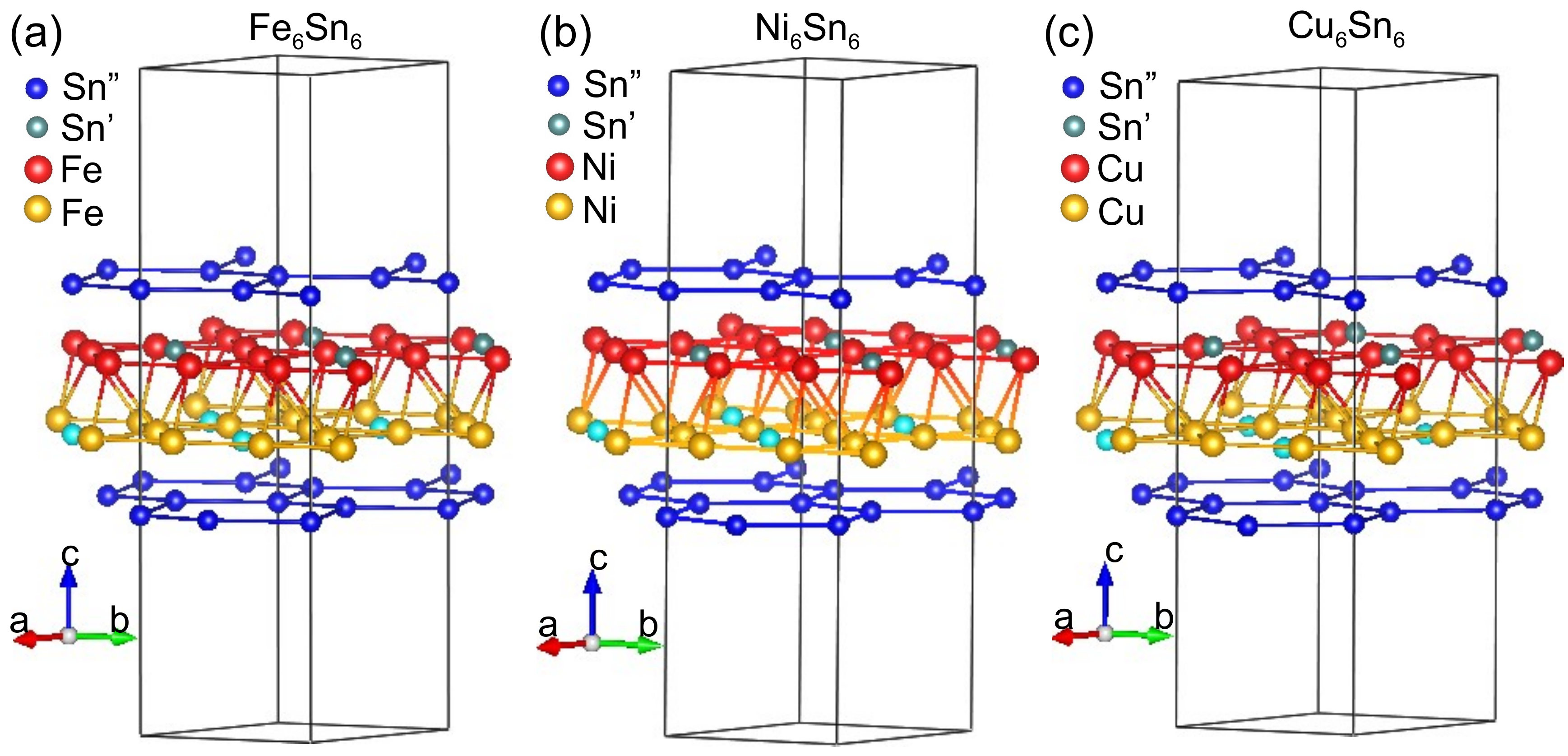}
\caption{Structure of bilayer (BL) (a) Fe$_6$Sn$_6$ , (b) Ni$_6$Sn$_6$ and (c) Cu$_6$Sn$_6$ compounds.}
\label{fig:bl_struct}
\end{figure*}

The fully relaxed crystal structure parameters of the bulk Fe$_{3}$Sn$_{2}$, Ni$_{3}$Sn$_{2}$ and Cu$_{3}$Sn$_{2}$ and the corresponding bilayers
are tabulated in the \Tab{tab:bulk_struct} and \Tab{tab:bl_struct}, respectively. To aid visualization, the structures are also presented in \Fig{fig:bulk_struct} and \Fig{fig:bl_struct}, respectively. In the bulk geometry, the metal
ions form bilayers of kagome networks, consisting of big and small equilateral triangles, Sn$'$ atoms occupy the position
close to the center of the hexagons in the kagome planes, while the kagome bilayers are separated by Sn$''$ layers.
The optimized geometries show that for Ni and Cu compounds, the Sn$'$ atoms are pushed further towards the direction of
kagome plane in the case of Ni and away in the case of Cu, compared to that of Fe. The difference in metal-metal bond length, d$^{M-M}$ between
the two triangles of the metal cations, was found to decrease with increase in atomic number from Fe to Cu. 
The difference between big and small triangles were found to be diminished for Ni compounds, with metal-metal distances of 2.73 $\AA$ and 2.55 $\AA$, compared to that of Fe compounds (2.78 $\AA$ and 2.54 $\AA$) while for Cu compound it was found to be vanishing (2.79 $\AA$ and 2.77 $\AA$) (cf. \Tab{tab:bulk_struct}). The difference of  d$^{M-M}$ between big and small triangles was found to be marginally
effected in moving from 3D to bilayer geometries.

\begin{table}[htb]
\resizebox{\columnwidth}{!}{%
\begin{tabular}{lllllll}
\hline
\multicolumn{6}{l}{Fe$_{3}$Sn$_{2}$, space group=R-3m (166), a=b=5.328 \AA,c=19.791 \AA,}\\
\multicolumn{6}{l}{ $\beta$=120 $^{\circ}$} \\\hline
Sites                       & Wyckoff         & x                 & y                 & z               & d$^{M-M}$ (\AA)   \\
Fe                          & h               & 0.17416           & -0.17416          & 0.21958         & 2.78,2.54         \\
Sn$^{\prime}$                  & c               & 0.0               & 0.0               & 0.10536         &                   \\
Sn$^{\dprime}$                 & c               & 0.0               & 0.0               & 0.33054           
\\
\hline
\multicolumn{6}{l}{Ni$_{3}$Sn$_{2}$, space group=R-3m (166), a=5.283 \AA,c=19.715 \AA}\\
\multicolumn{6}{l}{ $\beta$=120 $^{\circ}$} \\\hline
Sites                       & Wyckoff         & x                 & y                 & z               & d$^{M-M}$ (\AA)   \\
Ni                          & h               & 0.17239           & -0.17239          & 0.22493         & 2.73,2.55         \\
Sn$^{\prime}$                  & c               & 0.0               & 0.0               & 0.11595         &                   \\
Sn$^{\dprime}$                 & c               & 0.0               & 0.0               & 0.33005
\\
\hline
\multicolumn{6}{l}{Cu$_{3}$Sn$_{2}$, space group=R-3m (166), a=5.562 \AA,c=19.269 \AA }\\
\multicolumn{6}{l}{ $\beta$=120 $^{\circ}$} \\\hline        
Sites                       & Wyckoff         & x                 & y                 & z               & d$^{M-M}$ (\AA)   \\
Cu                          & h               & 0.16735           & -0.16735          & 0.21831         & 2.79,2.77         \\
Sn$^{\prime}$                  & c               & 0.0               & 0.0               & 0.09249         &                   \\
Sn$^{\dprime}$                 & c               & 0.0               & 0.0               & 0.33218
\\
\end{tabular}
}
\caption{Optimized geometry of bulk Fe$_{3}$Sn$_{2}$, Ni$_{3}$Sn$_{2}$ and Cu$_{3}$Sn$_{2}$}
\label{tab:bulk_struct}
\end{table}

\begin{table}[htb]
\resizebox{\columnwidth}{!}{%
\begin{tabular}{lllllll}
\hline
\multicolumn{6}{l}{Fe$_{6}$Sn$_{6}$, space group=P-3m1 (164), a=5.328 \AA,c=19.790 \AA}\\
\multicolumn{6}{l}{ $\beta$=120 $^{\circ}$} \\\hline
Sites                       & Wyckoff         & x                 & y                 & z               & d$^{M-M}$ (\AA)    \\
Fe                          & i               & -0.15916          & 0.15916           & 0.44747         & 2.784,2.544        \\
Sn$^{\prime}$                  & d               &  0.33333          & 0.66667           & 0.43689         &                    \\
Sn$^{\dprime}$                 & c               &  0.0              & 0.0               &-0.32606         &                    \\
Sn$^{\dprime}$                 & d               &  0.33333          & 0.66667           &-0.33309       
\\
\hline
\multicolumn{6}{l}{Ni$_{6}$Sn$_{6}$, space group=P-3m1 (164), a=5.283 \AA,c=19.715 \AA}\\
\multicolumn{6}{l}{ $\beta$=120 $^{\circ}$} \\\hline
Sites                       & Wyckoff         & x                 & y                 & z               & d$^{M-M}$ (\AA)     \\
Ni                          & i               & -0.16358          & 0.16358           & 0.44368         & 2.690,2.593         \\
Sn$^{\prime}$                  & d               &  0.33333          & 0.66667           & 0.44434         &                     \\
Sn$^{\dprime}$                 & c               &  0.0              & 0.0               &-0.33014         &                     \\
Sn$^{\dprime}$                 & d               &  0.33333          & 0.66667           &-0.33255    
\\
\hline
\multicolumn{6}{l}{Cu$_{6}$Sn$_{6}$, space group=P-3m1 (164), a=5.562 \AA,c=19.268 \AA}\\
\multicolumn{6}{l}{ $\beta$=120 $^{\circ}$} \\\hline        
Sites                       & Wyckoff         & x                 & y                 & z               & d$^{M-M}$ (\AA)     \\
Cu                          & i               & -0.16651          & 0.16651           & 0.44626         & 2.781,2.781         \\
Sn$^{\prime}$                  & d               &  0.33333          & 0.66667           & 0.42901         &                     \\
Sn$^{\dprime}$                 & c               &  0.0              & 0.0               &-0.32710         &                     \\
Sn$^{\dprime}$                 & d               &  0.33333          & 0.66667           &-0.32681        
\\
\end{tabular}
}
\caption{Optimized geometry of 2D bilayer Fe$_{6}$Sn$_{6}$, Ni$_{6}$Sn$_{6}$ and Cu$_{6}$Sn$_{6}$}
\label{tab:bl_struct}
\end{table}

\section{Analysis using model tight binding Hamiltonian}
\begin{figure*}
\centering
	\includegraphics[width=0.85\linewidth]{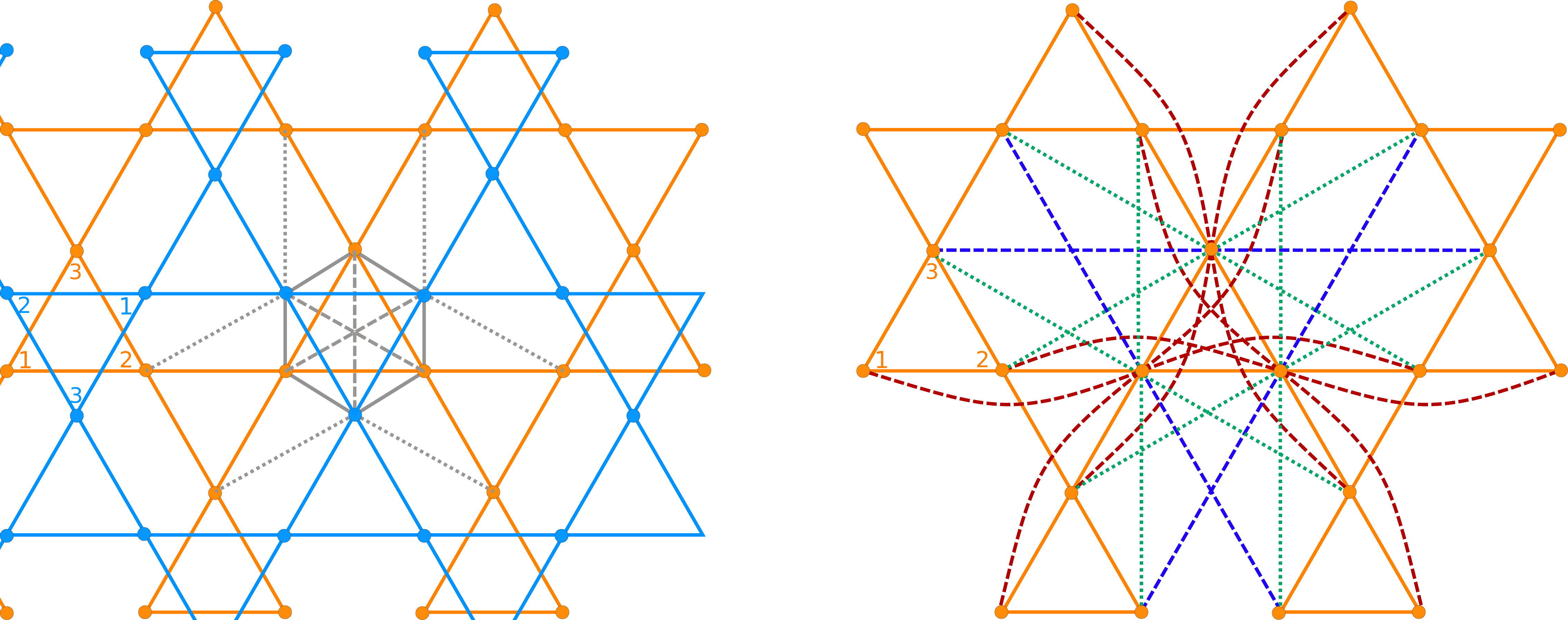}
	\caption{{\bf Hopping structure:} The hopping structure of each orbital species out-of-plane and in-plane are shown in the left and right diagrams, respectively. The unit cell is marked with indices in the left, where blue indices mark the sites in the top layer and the orange indices mark the sites in the bottom layer. {\it Left schematic} There is one out-of-plane nearest neighbour (nn) hopping $t_{\perp \chi}$ and it operates only within a unit cell (solid grey lines). There are two kinds of out-of-plane next nearest neighbour (nnn) hoppings, $t_{\perp 1 \chi}$ which operates within a unit cell (dashed grey lines) and $t_{\perp 1 \chi}'$ which operates between unit cells (dotted grey lines).
	{\it Right schematic} The nn hopping $t_\chi$ is shown by solid orange lines for the lower layer. There in only one variety of nnn hopping $t_{1\chi}$ which has been shown by dotted green lines. There are two varieties of next to next nearest neighbour (nnnn) hoppings. The first one operates along the solid orange lines $t_{2\chi}$ (dashed red lines) and the second one passes through the centre of the hexagons $t'_{2\chi}$ (dashed blue lines). The hopping structure in the upper layer can be obtained by rotating the lower plane by $180^\circ$ about the centre of an up triangle of the lower layer.}
	\label{fig:hopping_params}
\end{figure*}

To model the low energy band structure of bilayer Fe$_3$Sn$_2$ (Fig. 1(c) in the main text), following the DFT results, the following Fe spinful orbitals/site $\{d_{3z^2-r^2}^{(\text{Fe})}, d_{x^2-y^2}^{(\text{Fe})}, d_{xy}^{(\text{Fe})}\}$ were considered. Further, to take full advantage of the lattice symmetries, site dependent orientations of these orbitals, similar to Ref. \onlinecite{Xu15}, were defined. At the sites marked $1$ in a unit cell of \Fig{fig:hopping_params}
{\small
\begin{eqnarray}
|\tilde{d}_{x^2-y^2},1 \rangle & \equiv & \exp \left( \frac{-i}{\hbar}\hat{L}_z^{(1)} \frac{2\pi}{3} \right)|d_{x^2-y^2},1 \rangle \nonumber \\
& = & \cos \left( \frac{4\pi}{3} \right)|d_{x^2-y^2},1 \rangle + \sin \left( \frac{4\pi}{3} \right)|d_{xy},1 \rangle \nonumber \\
|\tilde{d}_{xy},1 \rangle & \equiv & \exp \left( \frac{-i}{\hbar}\hat{L}_z^{(1)} \frac{2\pi}{3} \right)|d_{xy},1 \rangle \nonumber \\
& = &\cos \left( \frac{4\pi}{3} \right)|d_{xy},1 \rangle - \sin \left( \frac{4\pi}{3} \right)|d_{x^2-y^2},1 \rangle 
\label{eq:d_at_1}
\end{eqnarray}}
while at sites marked by $2$ in \Fig{fig:hopping_params}
{\small
\begin{eqnarray}
|\tilde{d}_{x^2-y^2},2 \rangle & \equiv & \exp \left( \frac{-i}{\hbar}\hat{L}_z^{(2)} \frac{\pi}{3} \right)|d_{x^2-y^2},2 \rangle \nonumber \\
& = & \cos \left( \frac{2\pi}{3} \right)|d_{x^2-y^2},2 \rangle + \sin \left( \frac{2\pi}{3} \right)|d_{xy},2 \rangle \nonumber \\
|\tilde{d}_{xy},2 \rangle & \equiv & \exp \left( \frac{-i}{\hbar}\hat{L}_z^{(2)} \frac{\pi}{3} \right)|d_{xy},2 \rangle \nonumber \\
& = & \cos \left( \frac{2\pi}{3} \right)|d_{xy},2 \rangle - \sin \left( \frac{2\pi}{3} \right)|d_{x^2-y^2},2 \rangle \ . 
\label{eq:d_at_2}
\end{eqnarray}}
All the orbitals at the sites marked $3$ and the $d_{3z^2-r^2}$ orbitals at sites $1$ and $2$ remained the same as in the global basis.

The resultant tight-binding model had the generic form 
{\small
\begin{eqnarray}
	H = \sum_{\alpha \leq \beta} H_{\alpha \beta} & = & \sum _{\alpha \leq \beta} \left[ ( 1-\delta_{\alpha \beta}) (H_{\text{soc}}^{ \alpha \beta} + H_{nn}^{ \alpha\beta}) \right. \nonumber \\
	& + & \left. \delta_{\alpha \beta} (H_{\circ}^{ \alpha \beta} + H_{nn}^{ \alpha \beta} + H_{nnn}^{ \alpha \beta} + H_{nnnn}^{ \alpha \beta} ) \right]
	\label{eq:H_full}
\end{eqnarray}}
where $\alpha$ and $\beta$ denote the orbitals; with inter-orbital and intra-orbital terms made explicit.
In \eqn{eq:H_full} $H_{\circ}^{\alpha \beta}$ is the onsite term originating from the onsite energy (the kagome crystal field splits the $\tilde{d}_{3z^2-r^2}$ orbital from the $\tilde{d}_{x^2-y^2}$ and $\tilde{d}_{xy}$ orbitals; cf Ref. \onlinecite{Xu15}) and the background magnetization (arising from a mean field treatment of the onsite interaction term {\small $Un_{i\uparrow}n_{i\downarrow} = \frac{U}{2} \left[ n_{i\uparrow} + n_{i\downarrow} - \frac{1}{3} S_i^2\right] \simeq \frac{U}{2} \left[ n_{i\uparrow} + n_{i\downarrow} - S_i^z h\right] $
$= \frac{U}{2} \left[ (1-h) n_{i\uparrow} + (1+h) n_{i\downarrow} \right] $}; cf our DFT results). $H_{nn}^{\alpha \beta}$ is the collection of nearest neighbour hopping terms, wherein, to incorporate the breathing anisotropy between the small triangles and the big triangles (see Sec. \ref{sec:xtal_struct}) a parameter 
\begin{align}
r_{\alpha \beta} =t^{\text{out}}_{\alpha \beta}/t^{\text{in}}_{\alpha \beta}
\end{align}
was used. Here $t^{\text{in}}_{\alpha \beta}$ and $t^{\text{out}}_{\alpha \beta}$ are, respectively, the in-plane nearest neighbour hopping amplitudes between two orbitals $\alpha$ and $\beta$ when both the orbitals are within the same unit cell (see \Fig{fig:hopping_params}), and when they are in different unit cells. For the sake for simplicity $r_{\alpha \beta}=r$ was chosen to be independent of the orbital indices. $H_{nnn}^{\alpha \beta}$ and $H_{nnnn}^{\alpha \beta}$ denote the next-to-nearest-neighbour and next-to-next-nearest-neighbour hopping terms. The SOC term, though small, played an important role. It was obtained by projecting the atomic spin-orbit coupling to the above manifold of orbitals. The projected Hamiltonian had the following second quantized form
\begin{eqnarray}
	H_{\text{soc}}^{\alpha \beta} = i \lambda_{so} & \sum_{i,s,\gamma} & \left[ c^\dagger _{i3s\gamma \uparrow} c^{}_{i2s\gamma \uparrow} - c^\dagger _{i3s\gamma \downarrow} c^{}_{i2s\gamma \downarrow} \right. \nonumber \\
& & \left. - c^\dagger _{i2s\gamma \uparrow} c^{}_{i3s\gamma \uparrow} + c^\dagger _{i2s\gamma \downarrow} c^{}_{i3s\gamma \downarrow} \right]
	\label{eq:H_soc}
\end{eqnarray}
where the subscripts $i$, $\chi$, $s$, $\gamma$ and $\sigma$ indicate lattice site, orbital ($\chi \in \left\{ \tilde{d}_{3z^2-r^2}, \tilde{d}_{x^2-y^2}, \tilde{d}_{xy} \right\} \equiv \{1,2,3\}$), site index within a unit cell, layer number and spin, respectively.

Furthermore, the choice to work with the $\tilde{d}$ orbitals (\eqn{eq:d_at_1} and (\ref{eq:d_at_2})) led to certain constraints on the relative signs of the inter-orbital hopping amplitudes between the orbitals at different sites. These constraints for in-plane nearest neighbour hopping amplitudes, which were incorporated in our analysis, amount to having alternating signs of the hopping amplitudes between a $\tilde{d}_{x^2-y^2}$ (or $\tilde{d}_{3z^2-r^2}$) orbital and its four nearest neighbour $\tilde{d}_{xy}$ orbitals. For concreteness the explicit form of the inter-orbital nearest neighbour hopping terms inside the unit cell (defined in \Fig{fig:hopping_params}) are provided below.
{\small
\begin{eqnarray}
	- t_{12} \left[ (c_{i23\gamma\sigma}^\dagger + c_{i22\gamma\sigma}^\dagger) c_{i11\gamma\sigma} \right. & + & (c_{i21\gamma\sigma}^\dagger + c_{i23\gamma\sigma}^\dagger) c_{i12\gamma\sigma} \nonumber \\
	& + & \left. (c_{i22\gamma\sigma}^\dagger + c_{i21\gamma\sigma}^\dagger) c_{i13\gamma\sigma} + \text{h.c.} \right] \nonumber \\
	- t_{13} \left[ (c_{i13\gamma\sigma}^\dagger - c_{i12\gamma\sigma}^\dagger) c_{i31\gamma\sigma} \right. & + & (c_{i11\gamma\sigma}^\dagger - c_{i13\gamma\sigma}^\dagger) c_{i32\gamma\sigma} \nonumber \\
	& + & \left. (c_{i12\gamma\sigma}^\dagger - c_{i11\gamma\sigma}^\dagger) c_{i33\gamma\sigma} + \text{h.c.} \right] \nonumber \\
	- t_{23} \left[ (c_{i23\gamma\sigma}^\dagger - c_{i22\gamma\sigma}^\dagger) c_{i31\gamma\sigma} \right. & + & (c_{i21\gamma\sigma}^\dagger - c_{i23\gamma\sigma}^\dagger) c_{i32\gamma\sigma} \nonumber \\
	& + & \left. (c_{i22\gamma\sigma}^\dagger - c_{i21\gamma\sigma}^\dagger) c_{i33\gamma\sigma} + \text{h.c.} \right] \nonumber \\
	\label{<++>}
\end{eqnarray}}
where $t_{12}$, $t_{13}$ and $t_{23}$ are the in-plane nearest neighbour inter-orbital hopping amplitudes inside a unit cell between, respectively, a $\tilde{d}_{3z^2-r^2}$ orbital and a $\tilde{d}_{x^2-y^2}$ orbital, a $\tilde{d}_{3z^2-r^2}$ orbital and a $\tilde{d}_{xy}$ orbital, and a $\tilde{d}_{x^2-y^2}$ orbital and a $\tilde{d}_{xy}$ orbital. All the relevant parameters in our model and their meanings are shown schematically in \Fig{fig:hopping_params} and enumerated in Tables \ref{tab:lat_parameters}-\ref{tab:inter_parameters}.

\begin{table*}
\centering
\begin{tabular}{|c|c|c|}
	\hline
	parameters & description & value \\ \hline
	$\vec{a}_1$ & lattice vector & $(1,0)$ \\
	$\vec{a}_2$ & lattice vector & $(1/2,\sqrt{3}/2)$ \\ \hline
	$\vec{K}_1$ & reciprocal lattice vector & $2\pi (1, -1/\sqrt{3})$ \\
	$\vec{K}_2$ & reciprocal lattice vector & $2\pi (0, 2/\sqrt{3})$ \\ \hline
\end{tabular}
	\caption{{\bf Lattice parameters:} Lattice vectors and reciprocal lattice vectors.} \label{tab:lat_parameters}
\end{table*}

\begin{table*}
\centering
\begin{tabular}{|c|c|c|c|c|}
	\hline
	parameters & description & $\chi = 1$ & $\chi=2$ & $\chi = 3$ \\ \hline
	$t_{\chi}$ & nn in-plane hopping between Fe sites inside the unit cell & -0.07 & -0.405 & 0.185 \\
	$r_{\chi} = r$ & ratio between the nn hopping parameters outside the unit cell to those inside the unit cell & 1.25 & 1.25 & 1.25 \\
	$t_{\perp \chi}$ & nn out-of-plane hopping between Fe sites & 0.06 & -0.185 & 0.045 \\
	$t_{1 \chi}$ & nnn in-plane hopping between Fe sites & -0.03 & 0.075 & -0.135 \\
	$t_{\perp 1 \chi}$ & nnn out-of-plane hopping between Fe sites within a unit cell & 0.02 & 0.06 & -0.01 \\
	$t'_{\perp 1 \chi}$ & nnn out-of-plane hopping between Fe sites connecting different unit cells & 0.03 & -0.09 & -0.08 \\
	$t_{2 \chi}$ & nnnn in-plane hopping shown by dashed red lines in \Fig{fig:hopping_params} & -0.01 & -0.03 & 0.075 \\
	$t'_{2 \chi} $ & nnnn in-plane hopping shown by dashed blue lines in \Fig{fig:hopping_params} & 0.01 & 0.105 & -0.045 \\
	$h_\chi$ & background magnetization & 1.35 & 1.35 & 1.35 \\
	$\epsilon_\chi$ & onsite energy of the orbitals & 0.69 & 1.82 & 1.82 \\ \hline
\end{tabular}
\caption{{\bf Intra-orbital parameters:} Parameters connecting orbitals of the same flavor ($\chi$) including onsite and hopping terms (for details see \Fig{fig:hopping_params}). $\chi \in \{\tilde{d}_{3z^2-r^2},\tilde{d}_{x^2-y^2},\tilde{d}_{xy}\} \equiv \{1,2,3\}$.} \label{tab:intra_parameters}
\end{table*}

\begin{table*}
\centering
\begin{tabular}{|c|c|c|}
	\hline
	parameters & description & value \\ \hline
	$t_{12}$ & nn in-plane hopping between $\tilde{d}_{3z^2-r^2}$ and $\tilde{d}_{x^2-y^2}$ orbitals at Fe sites inside the unit cell & -0.04\\
	$r_{12} = r$ & ratio between the nn hopping between $\tilde{d}_{3z^2-r^2}$ and $\tilde{d}_{x^2-y^2}$ orbitals outside the unit cell to those inside the unit cell & 1.25 \\
	$t_{13}$ & nn in-plane hopping between $\tilde{d}_{3z^2-r^2}$ and $\tilde{d}_{xy}$ orbitals at Fe sites inside the unit cell & 0.05\\
	$r_{13} = r$ & ratio between the nn hopping between $\tilde{d}_{3z^2-r^2}$ and $\tilde{d}_{xy}$ orbitals outside the unit cell to those inside the unit cell & 1.25 \\
	$t_{23}$ & nn in-plane hopping between $\tilde{d}_{x^2-y^2}$ and $\tilde{d}_{xy}$ orbitals at Fe sites inside the unit cell & 0.14\\
	$r_{23} = r$ & ratio between the nn hopping between $\tilde{d}_{x^2-y^2}$ and $\tilde{d}_{xy}$ orbitals outside the unit cell to those inside the unit cell & 1.25 \\
	$\lambda_{23}$ & strength of spin-orbit coupling which operates only between $\tilde{d}_{x^2-y^2}$ and $\tilde{d}_{xy}$ orbitals & 0.11\\ \hline
\end{tabular}
	\caption{{\bf Inter-orbital parameters:} Parameters connecting orbitals of different flavor, including the spin-orbit coupling term.} \label{tab:inter_parameters}
\end{table*}

\subsection{Computation of DC Hall conductivity}
The real part of DC Hall conductivity, $\sigma_{xy}$, within linear response theory is
\begin{equation}
\sigma_{x y} = i \sum_{n\neq0} \frac{\langle 0 | J_y | n \rangle \langle n | J_x |0\rangle - \langle 0 | J_x | n \rangle \langle n | J_y |0\rangle}{(E_n - E_0)^2}
\end{equation}
where $E_{n(0)}$ is the energy of the eigenstate $|n\rangle$ ($|0\rangle\equiv$ ground state) of the system and $J_\mu$ is the total current operator 
\begin{equation}
J_\mu = \frac{i}{2} \sum_{j,\vec{\delta},\alpha,\beta} \delta_\mu t_{\alpha\beta}(\vec{\delta}) c_{j+\vec{\delta} \alpha}^\dagger c_{j\beta}^{} + h.c.
\end{equation}
Here $\alpha$ ($\beta$) is a collective index standing for $\tilde{d}$-orbital index, site index inside a unit cell, layer index and spin index, simultaneously. The hopping Hamiltonian characterized by the hopping amplitudes $t_{\alpha \beta} (\vec{\delta})$ between an orbital $\beta$ in a unit cell to an orbital $\alpha$ in another unit cell separated by the lattice vector $\vec{\delta}$, was diagonalized to compute the Hall conductivity for our tight binding (TB) model. The result was plotted in Fig. 4(b) of the main text as a function of the breathing anisotropy parameter $r$.

\subsection{Superconductivity\label{subsec:SC}} 
\begin{figure}[!htp]
\centering
	\includegraphics[width=1.0\linewidth]{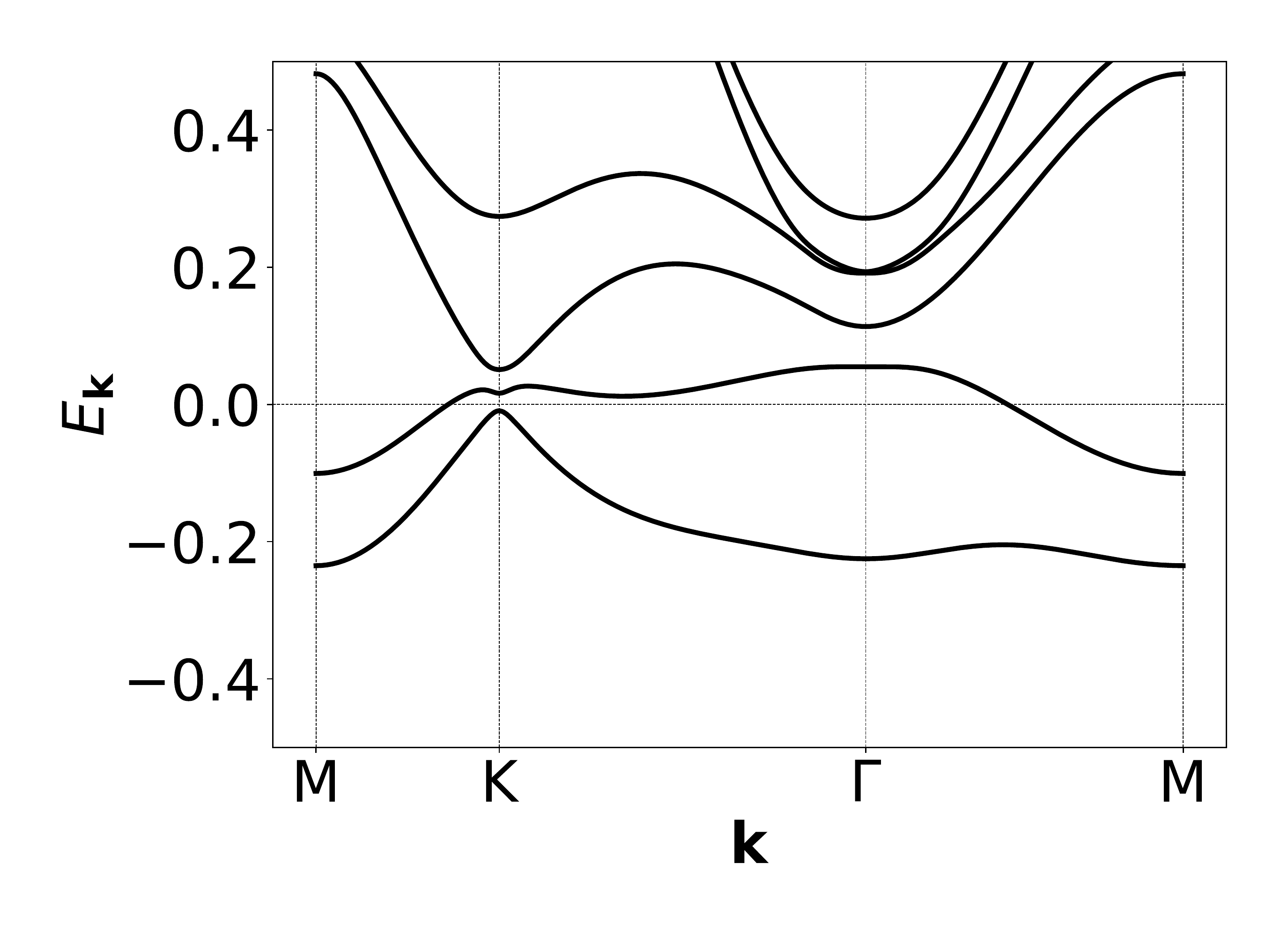}
	\caption{The tight binding bands obtained for parameters presented in Table \ref{tab:oneorbital_intra_parameters}.}
	\label{fig:band_reduced}
\end{figure}

As discussed in the main text, to capture the low energy superconducting instability of the Fermi surface, a one spin-polarised orbital/site TB model given by \eqn{eq:H_full} was used. The symmetry allowed effective SOC in this reduced model is given by $\tilde{H}_{\text{soc}} = \sum_{\langle I,J \rangle,\gamma} i\lambda \nu_{IJ}^{} c_{I \gamma \uparrow}^\dagger c_{J \gamma \uparrow}$, where $\lambda$ is the effective spin-orbit coupling parameter, $\gamma$ is the layer index and $\nu_{IJ}^{} = +1(-1)$ if $J \rightarrow I$ is anti-clockwise (clockwise) around the triangle containing the sites $I$ and $J$ (cf Ref. \onlinecite{Xu15}). Note that here the capital letter indices stand for the tuple $I\equiv(i,s)$ where $i$ is the unit cell index and $s$ is the site index inside the unit cell (cf \eqn{eq:H_soc}). In $\tilde{H}_{\text{soc}}$ only in-plane nearest neighbour $\langle I, J\rangle$ contribute. The parameters for this reduced model, presented in \Tab{tab:oneorbital_intra_parameters}, were chosen so that the qualitative features of the DFT bands close to the Fermi energy are adequately represented (see \Fig{fig:band_reduced}). Note that as in the case of flat band arising in kagome lattice problem with only nearest neighbour hopping\cite{Guo09}, the flat band near the Fermi energy in \Fig{fig:band_reduced} arises from an intricate (partial) destructive interference of the different hopping processes, including the inter-layer ones, present in our model.

\begin{table*}
\centering
\begin{tabular}{|c|c|c|c|c|c|c|c|c|c|c|c|}
	\hline
	parameters & $t$ & $r$ & $t_{\perp}$ & $t_1$ & $t_{\perp 1}$ & $t'_{\perp 1}$ & $t_2$ & $t'_2$ & $h$ & $\epsilon$ & $\lambda$ \\ \hline
	values & 0.13 & 1.25 & 0.06 & -0.1 & 0.02 & 0.0 & 0.05 & 0.01 & 1.35 & 1.67 & 0.01 \\ \hline
\end{tabular}
	\caption{{\bf Parameters for reduced TB model:} Parameters of our one-orbital / site model which capture important features of the DFT bands close to $E_{F}$.} \label{tab:oneorbital_intra_parameters}
\end{table*}

\begin{figure*}
	\centering
	\subfigure[]{
	\includegraphics[scale=0.4]{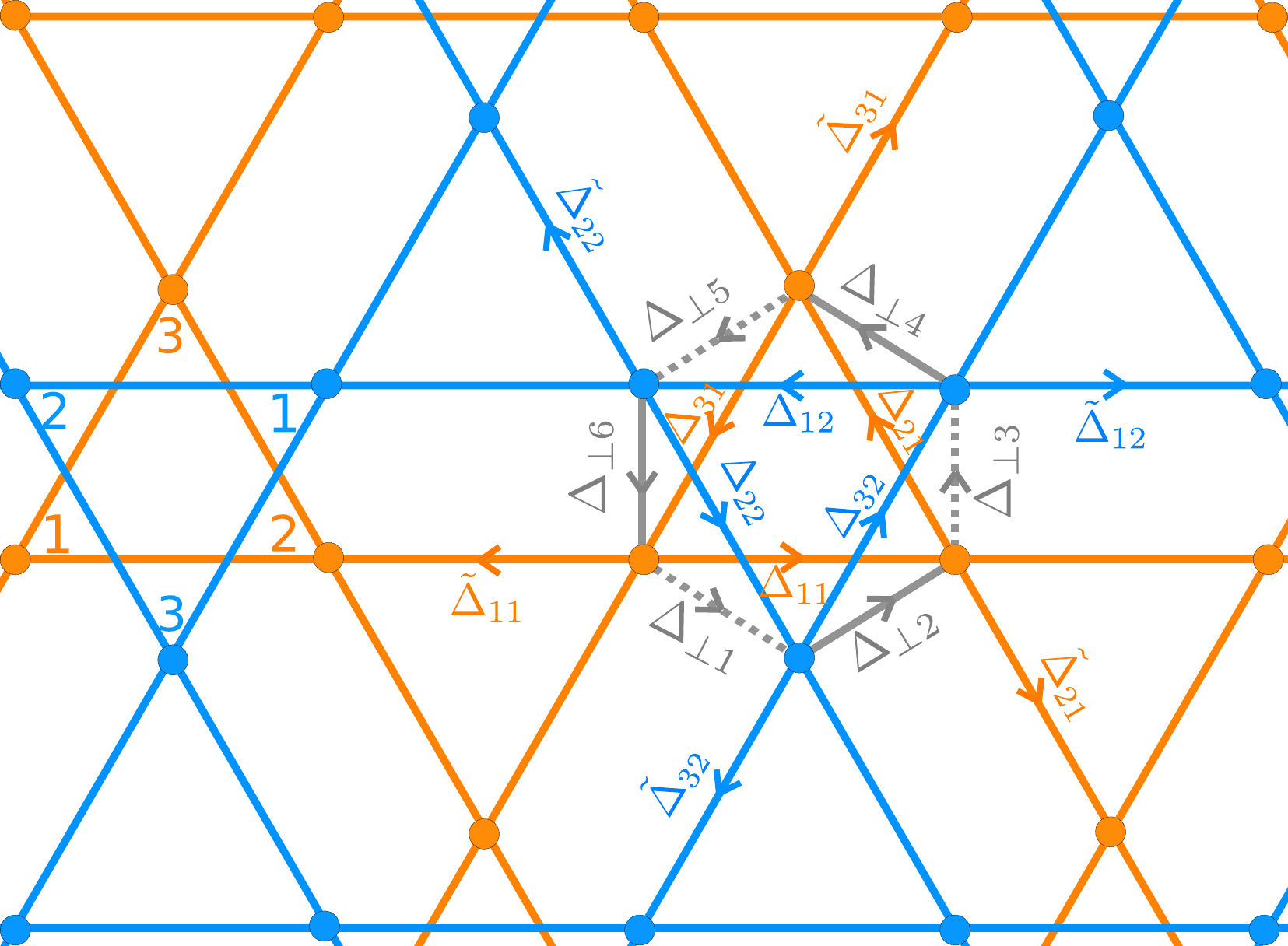}
	\label{fig:pairings}
	}\hspace*{0.5cm}
	\subfigure[]{
	\includegraphics[scale=1.5]{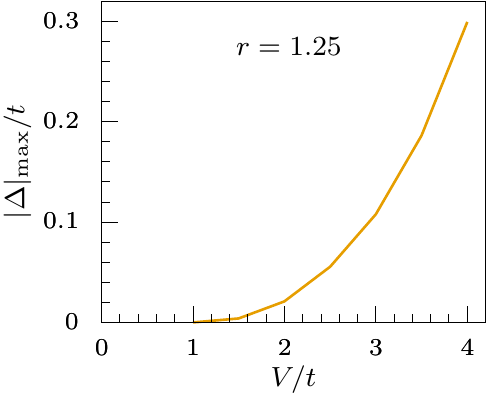}
	\label{fig:D_vs_V_1}
	}
	\caption{(a) The eighteen pairing amplitudes associated with an unit cell (marked by indices in the left side of the figure) are shown. The arrows on the bonds show the ordering of the fermionic operators used in the definition of the pairing amplitudes. For example $\Delta_{\perp 1} = \frac{V}{N} \sum_i \langle c_{i11\uparrow}^\dagger c_{i32\uparrow}^\dagger \rangle$. (b) Magnitude of the largest pair amplitude ($|\Delta|_{\text{max}} = |\Delta_{\perp 1}|$) as a function of the nearest neighbour attractive interaction $V$ in units of $t$ for tight binding parameters given in Table \ref{tab:oneorbital_intra_parameters}.}
\end{figure*}

\begin{figure*}
\centering
	\includegraphics[width=1.0\linewidth]{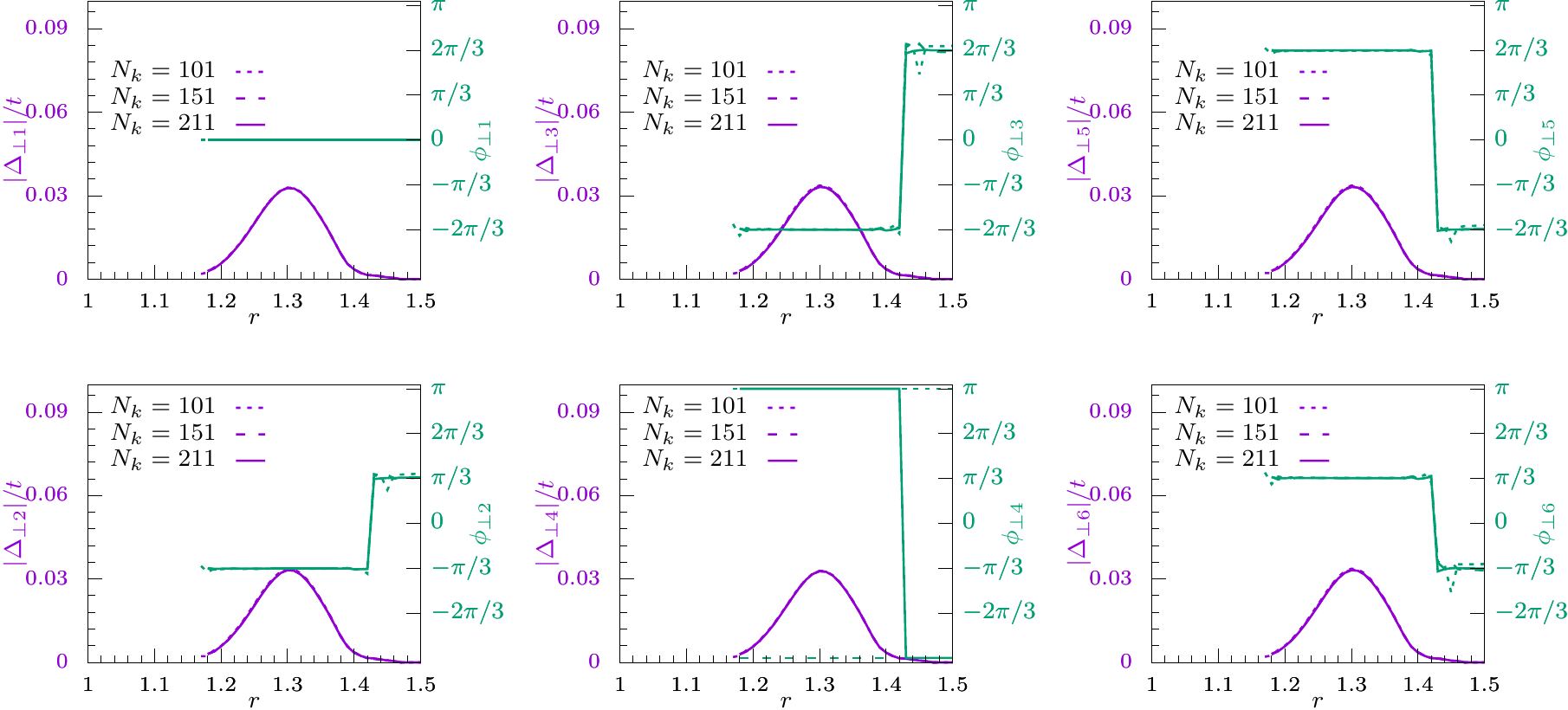}
	\caption{The out-of-plane SC order parameters (dominant bonds in Fig 4(c) of main text, i.e., $\Delta_{\perp 1}$, $\Delta_{\perp 2}$, $\Delta_{\perp 3}$, $\Delta_{\perp 4}$, $\Delta_{\perp 5}$, $\Delta_{\perp 6}$) have been plotted as a function of $r$. The amplitudes of the order parameters have been plotted on the left axis while their phases have been plotted on the right axis. Every other parameter is the same as in Table \ref{tab:oneorbital_intra_parameters}. Clearly, the results are robust to the different discretizations ($N_k \times N_k$) of the BZ used.}
	\label{fig:SC_vs_r}
\end{figure*}

An effective attractive interaction ($V$) on the nearest neighbour bonds, possibly arising because of spin-fluctuations\cite{Maier07}, was then introduced and the most generic mean-field superconducting state which does not break lattice translation symmetry was solved for. This state has eighteen variational parameters (pairing amplitudes) shown in Figure {\color{red}{S}}\ref{fig:pairings}.
The mean field Hamiltonian thus obtained, $\tilde{H}_{\text{SC}}$, and can be put in the Nambu form
\begin{widetext}
\begin{equation}
\tilde{H}_{\text{SC}} = \sum_{\veck>0} \psi_\veck^\dagger \mathcal{H}_\veck \psi_\veck \ \ \text{ where }\ \ \psi_\veck^\dagger \equiv \left[c_{\veck11\uparrow}^\dagger \ c_{\veck21\uparrow}^\dagger \ c_{\veck31\uparrow}^\dagger \ c_{\veck12\uparrow}^\dagger \ c_{\veck22\uparrow}^\dagger \ c_{\veck32\uparrow}^\dagger \  c_{-\veck32\uparrow} \ c_{-\veck22\uparrow} \ c_{-\veck12\uparrow} \ c_{-\veck31\uparrow} \ c_{-\veck21\uparrow} \ c_{-\veck11\uparrow} \right]
\end{equation}
\end{widetext}
where $\{c_{\veck s \gamma \sigma}\}$ are the Fourier components of $c_{i \chi s \gamma \sigma}$ in \eqn{eq:H_soc} with the orbital index $\chi$ dropped. $\tilde{H}_{SC}$ was then diagonalized using Bogoliubov transformation and all the pairing amplitudes were determined self-consistently.  In Fig. {\color{red}{S}}\ref{fig:D_vs_V_1} the maximum among the magnitude of the pairing amplitudes ($|\Delta|_{\text{max}} = |\Delta_{\perp 1}|$; see Fig. {\color{red}{S}}\ref{fig:pairings} and Fig. 4(c) of the main text) as a function of the nearest neighbour attraction $V$ has been plotted. Following the BCS nature of the self-consistent pairing amplitude equations, superconductivity set in when $V$ $\approx$ $t$.
    The dominant pairing amplitudes for $V=2t$ are shown in \Fig{fig:SC_vs_r} as a function of the breathing anisotropy parameter, $r$.


\subsection{Computing Chern number for the Bogoliubov bands}
With the self consistent pairing amplitudes, the Chern numbers of the negative energy Bogoliubov bands were calculated and their sum was used to characterize the topological nature of the mean field superconducting state of our system\cite{Tanaka12}. This sum was found to be odd for the SC that our system hosts, thus establishing its non-trivial topological nature (see Fig 4(d) of the main text).


\begin{thebibliography}{99}
\bibitem{footnote1}  The SOC calculations for Fe$_3$Sn$_2$ were carried out considering the magnetisation axis pointed along $x$- (M$_x$), $y$- (M$_y$) or $z$- (M$_z$) directions, which showed the energy corresponding to M$_z$ to be lower compared to M$_x$ or M$_y$.
  The results are thus shown for only $M_z$ direction.
\end{thebibliography}
\end{document}